\documentclass[journal=jacsat,layout=twocolumn,manuscript=article]{achemso}

\usepackage{lineno}
\usepackage{txfonts}
\usepackage[colorlinks=true, linkcolor=blue, urlcolor=blue, citecolor=blue]{hyperref}
\usepackage{braket}
\usepackage{pifont}
\newcommand{\cmark}{\ding{52}}
\newcommand{\xmark}{\ding{55}}

\title{How Symmetry Governs the Dihedral Angle Dependence of Intermolecular Spin-Orbit Coupling}

\author{Antonio J. Garz\'on-Ram\'irez}
\affiliation{Department of Chemistry, Institute for Quantum Information Research and Engineering, and Center for Molecular Quantum Transduction, Northwestern University, 2145 Sheridan Road, Evanston, Illinois 60208, USA}
\author{Connor K. Terry Weatherly}
\affiliation{Department of Chemistry, Institute for Quantum Information Research and Engineering, and Center for Molecular Quantum Transduction, Northwestern University, 2145 Sheridan Road, Evanston, Illinois 60208, USA}
\author{Kyle T. Kairys}
\affiliation{Department of Chemistry, Institute for Quantum Information Research and Engineering, and Center for Molecular Quantum Transduction, Northwestern University, 2145 Sheridan Road, Evanston, Illinois 60208, USA}
\author{Michael R. Wasielewski}
\affiliation{Department of Chemistry, Institute for Quantum Information Research and Engineering, and Center for Molecular Quantum Transduction, Northwestern University, 2145 Sheridan Road, Evanston, Illinois 60208, USA}
\author{Roel Tempelaar}
\affiliation{Department of Chemistry, Institute for Quantum Information Research and Engineering, and Center for Molecular Quantum Transduction, Northwestern University, 2145 Sheridan Road, Evanston, Illinois 60208, USA}
\email{roel.tempelaar@northwestern.edu}

\makeatletter
\let\l@addto@macro\relax
\makeatother
\usepackage[fontsize=11pt]{scrextend}

\begin{document}

\maketitle

\begin{abstract}
Spin-orbit, charge-transfer intersystem crossing (SOCT-ISC) allows for the efficient production of triplet excited states in donor-acceptor (DA) dyads without the involvement of heavy atoms, for use in a myriad of technologies. This process is commonly believed to proceed optimally when the donor and acceptor moieties are oriented under an orthogonal dihedral angle. Here, we challenge this idea through a theoretical study unveiling an orthogonal scenario where spin-orbit couplings (SOCs) are instead minimized. Such is rationalized based on an analysis of the structure-imposed symmetry properties of the involved singlet and triplet states. Notably, in this scenario, finite SOCs demand oblique orientation angles, which in turn requires molecular chirality, suggesting chirality to be a prerequisite for activating the involved SOC pathways.
\end{abstract}


\section{Introduction}

Triplet photosensitizers \cite{zhao_triplet_2013} are useful to many technological applications including photocatalysis \cite{shi_photoredox_2012,xuan_visiblelight_2012,hari_photocatalyzed_2013,hari_synthetic_2014,schultz_solar_2014,gunderson_photoinduced_2011}, light-emitting diodes \cite{baldo_highly_1998}, photovoltaics \cite{dai_platinumiibisaryleneethynylene_2012,congreve_external_2013,suzuki_photoinduced_2013}, photodynamic therapy \cite{clo_control_2007,celli_imaging_2010,awuah_boron_2012,kamkaew_bodipy_2013,stacey_new_2013,weijer_enhancing_2015,zhao_triplet_2015,li_activatable_2017}, and quantum information \cite{attwood_probing_2025,palmer_oriented_2024}. In recent years, spin-orbit, charge-transfer intersystem crossing (SOCT-ISC) has received great interest as a pathway for promoting triplet photosentisization in molecular donor-acceptor (DA) dyads \cite{zhao_recent_2018,hou_charge_2019,buck_spin-allowed_2019}. The overall SOCT-ISC process involves photoexcitation into a localized singlet excited state that separates into a charge-transfer state with singlet multiplicity, which is followed by recombination into a localized triplet state \cite{okada_ultrafast_1981,van_willigen_time-resolved_1996,gould_intersystem_2003,dance_time-resolved_2006}. Under favorable conditions, the SOCT-ISC mechanism involves spin-orbit couplings (SOCs) that are orders of magnitude larger than those driving ordinary spin-orbit intersystem crossing between $(\pi,\pi^*)$ states \cite{buck_spin-allowed_2019}, enabling high triplet yields in organic molecules without introducing the complications that arise under heavy atom substitutions \cite{yogo_highly_2005,adarsh_tuning_2010,sabatini_intersystem_2011,wu_organic_2011,zhao_triplet_2015}.

It has been argued that SOCs become optimized for an orthogonal dihedral angle between the donor and acceptor moieties \cite{okada_ultrafast_1981, van_willigen_time-resolved_1996, dance_time-resolved_2006, wang_bodipyanthracene_2017, sartor_exploiting_2018, liu_revisit_2018,dong_torsion-induced_2021, aster_long-lived_2021}, based on a loose analogy to atomic orbital based arguments \cite{el-sayed_spinorbit_1963,el-sayed_triplet_1968}. Accordingly, the molecular orbitals of the DA dyad being perpendicular creates sufficient orbital angular momentum to induce a change in spin angular momentum upon electron transfer. Yet, systematic investigations of the dihedral angle dependence of SOCT-ISC have remained lacking. Ultimately, the dihedral angle is governed by the structure and sterics of the molecular system and its surroundings \cite{van_willigen_time-resolved_1996}. As a result, studies have effectively compared angle variations across \emph{different} dyads in order to rationalize changes in SOCs \cite{okada_ultrafast_1981,van_willigen_time-resolved_1996,buck_spin-allowed_2019}. Among those studies, lack of a clear angle dependence \cite{buck_spin-allowed_2019} as well as efficient SOCT-ISC under non-orthogonal angles have been reported \cite{hou_electronic_2020,rehmat_carbazole-perylenebisimide_2020,xu_high-yield_2024,williams_mechanism_2022,williams_molecular_2025}. It remains unclear, however, whether these trends are solely attributable to the relative orientation of the donor and acceptor moieties, or also to variations in the electronic structure across the different dyads. Altogether, this leaves us without a clear picture of how SOCT-ISC can be optimized through tuning the dihedral angle.

Here, we report a computational study of the dihedral-angle dependence of SOCs in donor--acceptor dyads. Through constrained geometry optimizations, we systematically vary the dihedral angle between donor and acceptor dyads. We furthermore present results for which donor and acceptor moieties are separately geometry-optimized, whereupon their nuclear geometries are fused into oriented dyads. Through this route, and by separating out SOC contributions for different polarizations of the triplet spin configuration, we demonstrate that an orthogonal orientation can either enhance \emph{or suppress} the SOCT-ISC mechanism. Results are rationalized by analyzing the structure-imposed symmetries of the involved singlet and triplet states, showing that SOCs can rigorously vanish under orthogonal conditions. These findings open up a framework for the optimization of triplet photosensitizers.

\section{Results and Discussion}

\begin{figure}[t]
\includegraphics{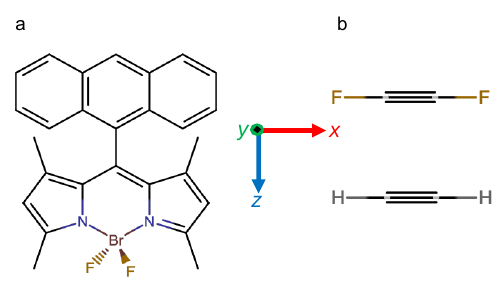}
\caption{Molecular structure of the dyads studied in this work, BD-anthracene (a) and the idealized system C$_2$H$_2$--C$_2$F$_2$ (b).}
\label{fig:structures}
\end{figure}

\begin{figure}[b!]
\includegraphics{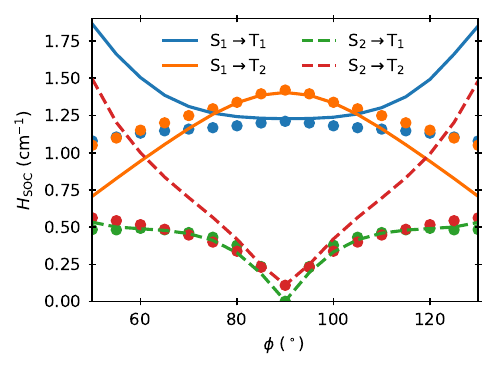}
\caption{Total SOCs between the two lowest-lying singlet excited states and triplet states as a function of dihedral angle for the BD-anthracene dyad. Shown are results obtained through a constrained geometry optimization of the dyad (markers) and through fusing moieties that were separately geometry-optimized (curves).}
\label{fig:total_SOC}
\end{figure}

In our analysis, we pay particular attention to the dyad consisting of boron dipyrromethene (BODIPY or BD) as the donor moiety fused to anthracene as the acceptor moiety, the molecular structure of which is shown in Fig.~\ref{fig:structures}a. The SOCT-ISC mechanism of this dyad was recently studied experimentally by Mani and coworkers \cite{buck_spin-allowed_2019}. The two lowest-lying singlet states of the dyad ($\mathrm{S}_1$ and $\mathrm{S}_2$) are known to be separated by only $\sim 0.15$ eV \cite{buck_spin-allowed_2019}, and for that reason we include both states in our analysis. Fig.~\ref{fig:total_SOC} shows calculated total SOCs associated with transitions from $\mathrm{S}_1$ and $\mathrm{S}_2$ to the lowest two triplets ($\mathrm{T}_1$ and $\mathrm{T}_2$), both of which are accessible from either singlet state. The total SOC between the $n$th singlet state and the $m$th triplet state is given by
\begin{equation}
  H_\mathrm{SOC} = \sqrt{\sum_{k=x,y,z}\left\vert\braket{\mathrm{S}_n|\hat{H}_\mathrm{SOC}|\mathrm{T}_{m,k}}\right\vert^2},
  \label{eq:total_SOC}
\end{equation}
where $\hat{H}_\mathrm{SOC}$ is the spin--orbit coupling Hamiltonian, and $\mathrm{T}_{m,k}$ features the index $k$ labeling three orthogonal spin-polarization components of the triplet manifold. Here and henceforth, reported calculations are conducted using PySCF \cite{sun_p_2018,sun_recent_2020}, and invoke time-dependent density functional theory (TD-DFT) within the Tamm-Dancoff approximation, using the PBE50 functional and the 6-311++G** basis set. As seen in Fig.~\ref{fig:total_SOC}, SOCs for each transition are sizable, with values of $\sim1$~cm$^{-1}$, in reasonable agreement with those reported for a fixed $\phi$ configuration by Mani and coworkers \cite{buck_spin-allowed_2019}.

The SOCs shown in Fig.~\ref{fig:total_SOC} are resolved as a function of the dihedral angle, $\phi$, between the BD and anthracene moieties. In common computational approaches, the dihedral angle is subject to the geometry optimization of the overal molecular structure \cite{buck_spin-allowed_2019}. As mentioned, we followed two approaches in order to exert dihedral angle control. In the first approach, a constrained geometry optimization was performed whereby we exclusively controlled the dihedral angle between the donor and acceptor moieties, $\phi$. In the second approach, the local nuclear configuration of each moiety was obtained from separate geometry optimizations. These configurations were then fused with a bond length of 1.48~\AA\ and held fixed while the relative orientation of the moieties was varied via the dihedral angle. The latter, hereafter referred to as the Fused Moiety Approach (FMA), allows us to isolate \emph{trends} in angle-dependent SOCs while minimizing inevitable complications arising under a constrained geometry optimization due to variations in the local moieties structure. Notably, sterics of the dyad restricted the permitted dihedral angle values to $50^\circ<\phi<130^\circ$, even for the FMA. As seen in Fig.~\ref{fig:total_SOC}, the SOCs calculated within both approaches are in reasonable agreement, especially near $\phi=90^\circ$ where steric effects are minimal. (Reported calculations were conducted only for $\phi\leq90^\circ$, as those for $\phi>90^\circ$ follow trivially upon mirror reflection. All geometry optimizations were conducted in vacuum.)

For both approaches, SOCs are seen to vary significantly with $\phi$. Notably, the value for $\mathrm{S}_1\rightarrow\mathrm{T}_2$ maximizes at $\phi=90^\circ$, consistent with predictions made in the literature \cite{okada_ultrafast_1981,van_willigen_time-resolved_1996,dance_time-resolved_2006,wang_bodipyanthracene_2017,sartor_exploiting_2018,liu_revisit_2018,dong_torsion-induced_2021}. However, the three other transitions lack a similar trend, and instead are broadly seen to \emph{minimize} at $\phi=90^\circ$, with $\mathrm{S}_2\rightarrow\mathrm{T}_1$ even vanishing at this angle. These results are inconsistent with the preconception that SOCs are maximized under orthogonality \cite{okada_ultrafast_1981,van_willigen_time-resolved_1996,dance_time-resolved_2006,wang_bodipyanthracene_2017,sartor_exploiting_2018,liu_revisit_2018,dong_torsion-induced_2021}, and warrant further investigation.

There are two factors preventing a deeper investigation based on the BD-anthracene results in Fig.~\ref{fig:total_SOC}. First, the total SOCs given by Eq.~\ref{eq:total_SOC}, although commonly used in calculations of ISC rates, obscure trends manifested in the underlying spin polarizations with $k=x$, $y$, and $z$. Second, the limited range of $\phi$ permitted in the BD-anthracene dyad complicates a comprensive study of angle-dependent couplings, and prevents study of the zero-angle limit, which is of significance as a point of high symmetry.

We address both limitations by pursuing a polarization-resolved exploration of SOCs for an idealized model dyad. The system of interest is shown in Fig.~\ref{fig:structures}b and consists of two carbon dimers capped with hydrogens and fluorines, C$_2$H$_2$ and C$_2$F$_2$, respectively. To simplify this model dyad as much as possible, we subjected it to the FMA, by first conducting a separate geometry optimization of C$_2$H$_2$ and C$_2$F$_2$ in vacuum. The resulting configurations were then held fixed at a separation of 3.5 \AA\ along an axis normal to the principal axis of each moiety. Even though no covalent bonds are expected to form between the C$_2$H$_2$ and C$_2$F$_2$ moieties, holding them in close proximity induces bonding-like behavior of their excited states as if bonding occurs, and the dyad thus formed will be referred to as C$_2$H$_2$--C$_2$F$_2$. Notably, the fluorine substitution on one of the carbon dimers introduces a chemical potential difference between the moieties, which avoids resonances that inhibit directed charge transfer.

\begin{figure}
\includegraphics{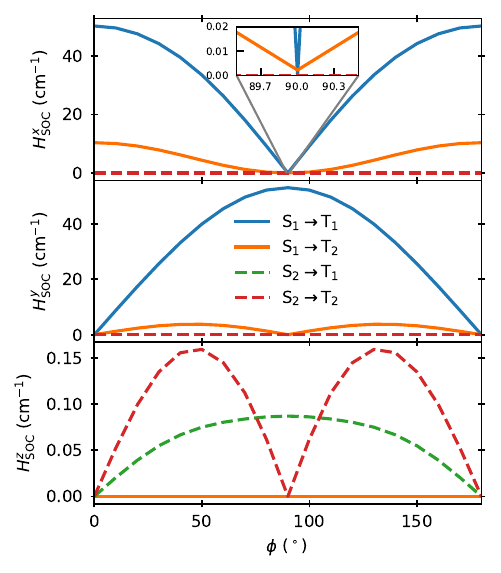}
\caption{Polarization-resolved SOCs for the two lowest-lying singlet excited states and triplet states as a function of dihedral angle for the idealized dyad C$_2$H$_2$--C$_2$F$_2$.}
\label{fig:carbon_dimers}
\end{figure}

Shown in Fig.~\ref{fig:carbon_dimers} are angle-resolved SOC magnitudes $\braket{\mathrm{S}_n\vert\hat{H}_\mathrm{SOC}\vert\mathrm{T}_{m,k}}$, for combinations of $\mathrm{S}_1$, $\mathrm{S}_2$, $\mathrm{T}_1$, and $\mathrm{T}_2$ in C$_2$H$_2$--C$_2$F$_2$, decomposed into polarization-resolved contributions. We define the polarization directions based on the molecular geometry (rather than based on the zero-field splitting tensor used to interpret magnetic resonance experiments). Specifically, $k=z$ corresponding to the direction along the axis connecting the two moieties, about which the dihedral angle is defined, and the $x$ direction is taken to lie in the plane of the dyad for $\phi=0^\circ$. Under this angle, we observe extraordinarily large SOCs for $k=x$, which are associated with electron gyration about the principal axes of the C$_2$H$_2$ and C$_2$F$_2$ moieties. Similar magnitudes are found for $k=y$ under finite angles, as principal axes are tilted out of the $xz$ plane. This effect will vanish for nonlinear moieties such as BD and anthracene. Instead, the $k=z$ component of C$_2$H$_2$--C$_2$F$_2$ involves values that are much more modest, reaching magnitudes in rough agreement with BD-anthracene. Regardless, all polarization-resolved SOCs show clean trends with extreme behaviors at $\phi=0^\circ$ and $\phi=90^\circ$ where SOC is maximized, minimized, and at times rigorously zero, broadly consistent with the observations made in Fig.~\ref{fig:total_SOC} for the total SOCs of BD-anthracene.

\begin{table*}[t!]
\caption{Dihedral-angle dependent irreps of the two lowest-lying singlet excited states and triplet states of C$_2$H$_2$--C$_2$F$_2$. Shown below are flags indicating whether or not SOCs are symmetry-allowed.}
\label{tab:carbon_dimers}
\begin{tabular}{c|ccc|ccc|ccc}
\hline \hline
  $\phi$ & & $0^\circ$ & & & $45^\circ$ & & & $90^\circ$ & \\ \hline \hline
  $k$&$x$&$y$&$z$&$x$&$y$&$z$&$x$&$y$&$z$ \\ \hline \hline
  $\mathrm{S}_1$&A$_2$&A$_2$&A$_2$&A&A&A&A$_2$&A$_2$&A$_2$\\ 
  $\mathrm{S}_2$&B$_1$&B$_1$&B$_1$&B&B&B&B$_1$&B$_1$&B$_1$\\
  $\mathrm{T}_{1,k}$&A$_2$&A$_1$&B$_2$&A&A&B&A$_1$&A$_2$&B$_1$ \\
  $\mathrm{T}_{2,k}$&A$_2$&A$_1$&B$_2$&A&A&B&A$_2$&A$_1$&B$_2$\\ \hline\hline
  $\mathrm{S}_1\rightarrow\mathrm{T}_{1,k}$ & \cmark & \xmark & \xmark & \cmark & \cmark & \xmark & \xmark & \cmark & \xmark \\

  $\mathrm{S}_1\rightarrow\mathrm{T}_{2,k}$ & \cmark & \xmark & \xmark & \cmark & \cmark & \xmark & \cmark & \xmark & \xmark \\

  $\mathrm{S}_2\rightarrow\mathrm{T}_{1,k}$  & \xmark & \xmark & \xmark & \xmark & \xmark & \cmark & \xmark & \xmark & \cmark \\

  $\mathrm{S}_2\rightarrow\mathrm{T}_{2,k}$  & \xmark & \xmark & \xmark & \xmark & \xmark & \cmark & \xmark & \xmark & \xmark \\ \hline\hline
\end{tabular}
\end{table*}

Even for the idealized model system C$_2$H$_2$--C$_2$F$_2$, the SOC values are nontrivially governed by many-body electronic interactions. It is therefore difficult to rationalize the SOC values at a quantitative level. However, a qualitative assessment can be conducted based on the symmetry properties of the involved singlet and triplet states. Notably, within the FMA we arrive at molecular structures with idealized symmetry properties, which translates to highly-symmetric electronic excitations. This permits a symmetry analysis based on which to establish principles for intermolecular SOC in analogy to the heuristic atomic orbital rules for intramolecular SOC proposed by El-Sayed \cite{el-sayed_spinorbit_1963}. Specifically, for SOC to be nonzero, $\braket{\mathrm{S}_n|\hat{H}_\mathrm{SOC}|\mathrm{T}_{m,k}}\neq0$, it is required that the tensor product of the irreducible representations (irreps) of the involved states $\mathrm{S}_n$ and $\mathrm{T}_{m,k}$, and the SOC Hamiltonian, $\hat{H}_\mathrm{SOC}$, is contained in the totally-symmetric irrep ($\Gamma_\mathrm{s}$) of the point group of the dyad. That is \cite{wigner_pure_1959, lipkowitz_spinorbit_2001},
\begin{equation}
  \Gamma(\mathrm{S}_n)\otimes \Gamma(\hat{H}_{\mathrm{SOC}})\otimes \Gamma(\mathrm{T}_{m.k}) \supset\ \Gamma_{\mathrm{s}},
\label{eq:symm}
\end{equation}
where $\Gamma(i)$ denotes the irrep of quantity $i$. Given that $\Gamma(\hat{H}_\mathrm{SOC})=\Gamma_\mathrm{s}$, this is equivalent to $\Gamma(\mathrm{S}_n)\otimes \Gamma(\mathrm{T}_{m,k}) \supset \Gamma_\mathrm{s}$. We note that, in our analysis, $\mathrm{S}_n$ and $\mathrm{T}_{m,k}$ are taken to represent spin and spatial product states, and the operator $\hat{H}_{\mathrm{SOC}}$ is retained in the full spin-orbit basis.

The symmetry of the dyadic structure depends on the dihedral angle, for which three cases can be differentiated. Two singular cases correspond to $\phi=0^\circ$ and $90^\circ$, where the structure reaches distinct symmetries. This contrasts with the generic case of an oblique dihedral angle, $0<\phi<90^\circ$, for which the symmetry tends to be lower. (Here and henceforth, we will restrict the discussion to $0\leq\phi\leq90^\circ$ while noting that the domain $90\leq\phi\leq180^\circ$ follows trivially upon mirror reflection.) These principles impact the electronic state symmetry, as becomes obvious in Table \ref{tab:carbon_dimers}, which shows the Mulliken symbols for the irreps of the two lowest-lying singlet excited states and triplet states of C$_2$H$_2$--C$_2$F$_2$ for $\phi=0^\circ$ and $90^\circ$, as well as an oblique angle value $\phi=45^\circ$. The irreps allow us to assess whether or not SOCs are allowed by symmetry for combinations of the states involved, as established by Eq.~\ref{eq:symm}.

Included in Table \ref{tab:carbon_dimers} are flags indicating whether or not SOCs are symmetry-allowed. Critically, instances where this analysis predicts SOCs to be symmetry-forbidden all correspond to vanishing coupling values in Fig.~\ref{fig:carbon_dimers}. Furthermore, this figure shows that coupling values typically undergo monotonic growth towards extrema across regions where SOCs are symmetry-allowed. Notably, there are also transitions for which SOCs are forbidden throughout all $\phi$ values, which is reflected both in Fig.~\ref{fig:carbon_dimers} and Table \ref{tab:carbon_dimers}. As such, we find the symmetry analysis to account for the behavior of the spin-polarization resolved SOCs.

\begin{table}[b!]
\caption{Dihedral-angle dependent irreps for the two lowest-lying singlet excited states and triplet states of the BD-anthracene dyad. Shown below are flags indicating whether or not SOCs are symmetry-allowed.}
\label{tab:BD_anthracene}
\begin{tabular}{c|ccc|ccc}
\hline\hline
$\phi$ & & $50^\circ$ & & & $90^\circ$ & \\ \hline \hline
$k$ & $x$ &$y$&$z$&$x$&$y$&$z$ \\ \hline \hline
$\mathrm{S}_1$&A&A&A&A$_2$&A$_2$&A$_2$\\ 
$\mathrm{S}_2$&B&B&B&B$_2$&B$_2$&B$_2$\\
$\mathrm{T}_{1,k}$&A&A&B&A$_1$&A$_2$&B$_1$ \\
$\mathrm{T}_{2,k}$&B&B&A&B$_2$&B$_1$&A$_2$\\ \hline\hline

$\mathrm{S}_1\rightarrow\mathrm{T}_{1,k}$ & \cmark & \cmark & \xmark &  \xmark & \cmark & \xmark \\

$\mathrm{S}_1\rightarrow\mathrm{T}_{2,k}$ & \xmark & \xmark & \cmark &  \xmark & \xmark & \cmark \\

$\mathrm{S}_2\rightarrow\mathrm{T}_{1,k}$ & \xmark & \xmark & \cmark &  \xmark & \xmark & \xmark \\

$\mathrm{S}_2\rightarrow\mathrm{T}_{2,k}$ & \cmark & \cmark & \xmark & \cmark & \xmark & \xmark \\ \hline\hline
\end{tabular}
\end{table}

\begin{figure}[t]
\includegraphics{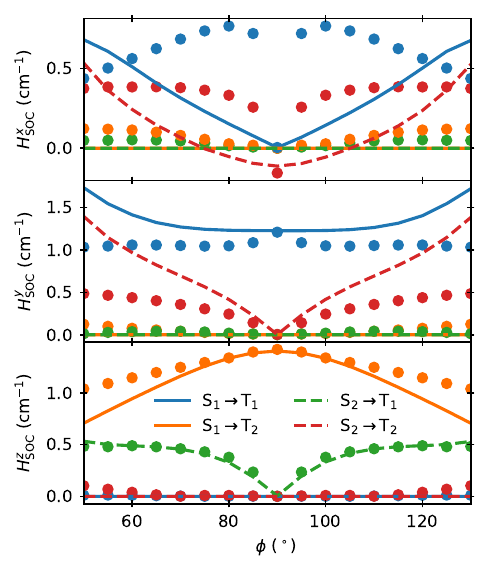}
\caption{As in Fig.~\ref{fig:total_SOC}, but for polarization-resolved SOCs.}
\label{fig:BD_anthracene}
\end{figure}

We proceed to apply a similar symmetry analysis to the BD-anthracene dyad, while defining the polarization directions analogously. (As with C$_2$H$_2$--C$_2$F$_2$, we first restrict ourselves to the FMA, since a constrained geometry optimization of the entire dyad entails subtle symmetry breaking, producing a low-symmetry conformation that does not permit a meaningful symmetry analysis.) Results are summarized in Table \ref{tab:BD_anthracene} for $\phi=50^\circ$ and $90^\circ$, which represent physically-relevant dihedral angle values permitted by the sterics of the dyad. As with C$_2$H$_2$--C$_2$F$_2$, we observe various combinations of transitions and angles where SOCs are symmetry-forbidden, which we expect to bear out in the SOC calculations. To assess this, we show in Fig.~\ref{fig:BD_anthracene} the polarization-resolved SOCs between the two lowest-lying singlet excited states and triplet states of BD-anthracene. Results are shown from the FMA as well as the constrained geometry optimization, the former of which should match the behavior predicted by Table \ref{tab:BD_anthracene}. Indeed, any instance marked as symmetry-forbidden in Table \ref{tab:BD_anthracene} manifest as a rigorously-vanishing value in the calculated SOCs. Notably, however, the $\mathrm{S}_2\rightarrow\mathrm{T}_{2,x}$ transition is seen to cross a zero value at an oblique angle of $\phi\sim 75 ^\circ$. This illustrates that the symmetry analysis provides a necessary but not sufficient criterion for SOCs to be nonzero. In Fig.~\ref{fig:BD_anthracene}, SOCs calculated under the constrained geometry optimization are seen to be in reasonable agreement with those from the FMA, with differences at small angles attributable to steric effects, and deviations close to perpendicular conditions being attributable to symmetry breaking arising under the constrained geometry optimization.

The total SOCs shown in Fig.~\ref{fig:total_SOC} follow from the polarization-resolved contributions from Fig.~\ref{fig:BD_anthracene} through substitution into Eq.~\ref{eq:total_SOC}, and the total SOCs can therefore be understood based on those contributions. In turn, any instance of nonvanishing contributions are rationalized by the symmetry analysis from Table \ref{tab:BD_anthracene}. Notably, the coupling associated with $\mathrm{S}_2\rightarrow\mathrm{T}_1$ is symmetry-forbidden at $\phi=90^\circ$ for all spin polarizations ($k=x$, $y$, and $z$). Similar behavior is observed for 
$\mathrm{S}_2\rightarrow\mathrm{T}_2$ in C$_2$H$_2$--C$_2$F$_2$. For this ideal model dyad, the lack of steric restrictions allows the dihedral angle to be scanned throughout, permitting us to make a remarkable observation: For the $\mathrm{S}_2\rightarrow\mathrm{T}_2$ SOC to be symmetry-allowed, the dyad needs to be oriented such that the dihedral angle is oblique.

Curiously, the oblique dihedral angle, required for certain singlet--triplet transitions to become symmetry-allowed, effectively renders the dyad chiral. It can thus be concluded that chirality is a fundamental requisite for the associated SOC to become activated. This in turns suggests a profound connection between spin effects and chirality, which resonates with the recent surge in studies also suggesting this connection \cite{ray_asymmetric_1999, naaman_chiral_2019, eckvahl_direct_2023}. Notably, the spin polarization for which SOC becomes activated aligns with the axis connecting the donor and acceptor moieties, about which the dihedral angle is defined. Altogether, this principle shows consistency with the experimental observation of increased $\mathrm{T}_z$ populations upon the introduction of chirality in donor--bridge--acceptor triads \cite{eckvahl_direct_2023}.

\section{Conclusion}

To conclude, we have explored SOCs between low-lying singlet and triplet states in two donor-acceptor dyads, with the aim to elucidate the principles driving SOCT-ISC. We have compared two approaches wherein geometry optimizations were applied to the entire dyad (while constraining the dihedral angle) versus the individual moieties, in order to systematically study dihedral angle dependence of SOCs. We have shown that spin-polarization resolved SOCs are well accounted for by a symmetry analysis, establishing a framework analogous to the El-Sayed rules, but at the intermolecular level. This framework could be helpful for ongoing efforts to engineer excited-state dynamics by tuning the dihedral angle in dyads \cite{estergreen_controlling_2022}. It rationalizes that SOCs could be maximized or minimized under an orthogonal angle, depending on the singlet and triplet state symmetries. It furthermore suggests a critical connection between SOCs and chirality for certain singlet--triplet transitions.

\section*{Acknowledgements}

This work was supported as part of the Center for Molecular Quantum Transduction, an Energy Frontier Research Center funded by the U.S.~Department of Energy, Office of Science, Basic Energy Sciences, under Award \#DE-SC0021314.

\bibliography{soct_isc_p}

\end{document}